\pdfoutput=1
\documentclass[a4paper, 10pt, conference]{IEEEtran}
\IEEEoverridecommandlockouts

\usepackage{geometry}
\usepackage{cite}
\usepackage{amsmath,amssymb,amsfonts}
\usepackage{amsmath}
\usepackage{graphicx}
\usepackage{textcomp}
\usepackage{xcolor}
\usepackage{algorithm}  
\usepackage{booktabs}
\usepackage{algpseudocode}  
\usepackage{amsmath}
\usepackage{threeparttable}
\usepackage{subfigure}

\usepackage[T1]{fontenc}
\usepackage{aecompl}

\usepackage{algorithm,algpseudocode}
\makeatletter
\newcommand{\algmargin}{\the\ALG@thistlm}
\makeatother
\newlength{\whilewidth}
\algdef{SE}[parWHILE]{parWhile}{EndparWhile}[1]
{\parbox[t]{\dimexpr\linewidth-\algmargin}{%
		\hangindent\whilewidth\strut\algorithmicwhile\ #1\ \algorithmicdo\strut}}{\algorithmicend\ \algorithmicwhile}%
\algnewcommand{\parState}[1]{\State%
	\parbox[t]{\dimexpr\linewidth-\algmargin}{\strut #1\strut}}

\def\BibTeX{{\rm B\kern-.05em{\sc i\kern-.025em b}\kern-.08em
		T\kern-.1667em\lower.7ex\hbox{E}\kern-.125emX}}
\begin{document}
	
	\title{A Multi-Modal States based Vehicle Descriptor and Dilated Convolutional Social Pooling for Vehicle Trajectory Prediction
		\thanks{}
	}
	
	\author{Huimin Zhang$^{1}$, Yafei Wang$^{1}$, Junjia Liu$^{1}$, Chengwei Li$^{1}$, Taiyuan Ma$^{1}$, Chengliang Yin$^{1}$
		\thanks{$^{1}$H. Zhang, Y. Wang, J. Liu, C. Li, T. Ma, C. Yin are with School of Mechanical Engineering,
			University of Shanghai Jiao Tong, Shanghai 200240, China.
		}%
	}

	\maketitle
	
	\begin{abstract}
		Precise trajectory prediction of surrounding vehicles is critical for decision-making of autonomous vehicles and learning-based approaches are well recognized for the robustness. However, state-of-the-art learning-based methods ignore 1) the feasibility of the vehicle's multi-modal state information for prediction and 2) the mutual exclusive relationship between the global traffic scene receptive fields and the local position resolution when modeling vehicles' interactions, which may influence prediction accuracy. Therefore, we propose a vehicle-descriptor based LSTM model with the dilated convolutional social pooling (VD+DCS-LSTM) to cope with the above issues. First, each vehicle's multi-modal state information is employed as our model's input and a new vehicle descriptor encoded by stacked sparse auto-encoders is proposed to reflect the deep interactive relationships between various states, achieving the optimal feature extraction and effective use of multi-modal inputs. Secondly, the LSTM encoder is used to encode the historical sequences composed of the vehicle descriptor and a novel dilated convolutional social pooling is proposed to improve modeling vehicles' spatial interactions. Thirdly, the LSTM decoder is used to predict the probability distribution of future trajectories based on maneuvers. The validity of the overall model was verified over the NGSIM US-101 and I-80 datasets and our method outperforms the latest benchmark.
		
	\end{abstract}

	\section{Introduction}
	During the development of autonomous driving, autonomous vehicles must go through complex traffic scenarios with vehicles driven by human drivers. In order to ensure that vehicles can drive safely and efficiently, autonomous vehicles must be capable of understanding the driving behavior of other vehicles and predicting the future trajectory of surrounding vehicles to achieve dynamic balance in traffic scenarios. Trajectory prediction is a process of inferring the future trajectory of a predicted vehicle based on historical information.
	
	In the early days, many researches adopted Bayesian formula \cite{bayes}, Monte Carlo simulation \cite{monte-carlo} and hidden Markov models (HMMs) \cite{HMM} such model-based methods for vehicle trajectory prediction. However, these methods are very sensitive to dynamic environments and it is difficult to obtain satisfactory accuracy in long-term prediction.
	In order to improve the predictive performance in the long time-domain, many researches have proposed learning-based methods. The trajectory sequence is a time sequence, therefore most of the researches\cite{social-lstm}\cite{imitate}\cite{cs-lstm}\cite{multi-modal}\cite{social-gan}\cite{Gail-gru} have used recurrent neural networks that are good at dealing with time sequences problems for trajectory prediction, which significantly enhances the predictive accuracy. 
	
	However, most of the recent learning-based researches only takes the trajectory positions of the vehicle as the model's input and do not utilize the vehicle's multi-modal state information. Obviously, this kind of input cannot accurately describe the vehicle's instantaneous motion states. A very intuitive idea is that the more detailed historical information we get, the more we can fully understand the surrounding traffic scene of the predicted vehicle, consequently making a more reasonable and accurate trajectory prediction.
	Actually, there are many factors that can influence driving behavior. The first category is the driver's own factors, including physical and psychological factors \cite{sarkar2019behavior}. The second category corresponds to traffic conditions, such as the type and size of surrounding vehicles \cite{jamson2013behavioural}. The third category of factors affecting driving behavior is related to the kinematics of the vehicle, such as position, speed, acceleration and heading angle, etc \cite{Semantic}.
	Therefore, we expand input information with the latter two factors in this article to make our multi-modal state input as detailed as possible. 
	While, when there are many features in the input and the interaction relationship between features is not clear, it is difficult for us to extract features artificially. 
	Simple features can be selected and extracted manually. However, the more abstract deep features cannot be filtered and extracted artificially, but can only be extracted automatically by an autoencoder\cite{zeiler2014visualizing}\cite{zhou2014stacked}.
	Du et al. \cite {du2016stacked} uses a stacked convolution denoising auto-encoder to automatically learn deep-level features from complex input images and improve the performance of subsequent support vector machine classifiers.
	Lei et al.\cite{lei2016intelligent} applied deep auto-encoders to directly learn features from signals for classification.
	Therefore, we introduce stacked sparse auto-encoders (SSAE) to automatically extract deep and high-level features from input data.
	
	In addition, all vehicles in the same traffic scene form an interdependent whole and their motions affect each other's decisions. Therefore, the spatial interaction between vehicles is very important for trajectory prediction. 
	When considering the interaction between vehicles, not only do we need to focus on the exact local interactive location but also we need to expand the receptive field of the global scene.
	Alahi et al.\cite{social-lstm} proposed social-LSTM, which applies social pooling to model the interaction between pedestrians. But this method doesn't directly utilize the spatial relationship between agents.
	Deo et al.\cite{cs-lstm} utilized the spatial relationship by convolutional layers, but used the max-pooling layer to reduce the resolution and expand the receptive field simultaneously. However, this method reduces the accuracy of local positions.
	It can be seen that these existing methods didn't solve the mutual exclusive problem between the global scene receptive field and the local position resolution when modeling interaction aware models.
	
	Recently, some researches have studied methods to deal with conflicts between large receptive fields and high-resolution prediction needs. Ronneberger et al. \cite{ronneberger2015u} proposed to use repeated upper convolutional layers to recover the lost resolution, while performing global perspective from the downsampling layer. The deconvolutional layer \cite{noh2015learning} can reduce the loss of information, but the additional complexity and execution delay may not be suitable for all situations. Yu et al. \cite{yu2015multi} has developed a new convolutional network module for dense prediction using dilated convolutions, without pooling or subsampling, which can aggregate large-scale context information without losing resolution, improving the accuracy of semantic segmentation. Zhen et al. \cite{zhen2019dilated} designed a dilated convolutional neural network architecture to increase the acceptance domain of each convolution.
	Consequently, we propose a novel dilated convolutional social pooling to extract interactive features in this paper.
	
	Finally, in this paper we propose a vehicle-descriptor based LSTM model with the dilated convolutional social pooling (VD+DCS-LSTM) for trajectory prediction in a highway scene.
	To summarize our main contributions as follows:
	\begin{itemize}
		\item[1.] We propose a novel \textit{vehicle descriptor} to achieve optimal feature representation of each vehicle's transient multi-modal state information. In this article, we expand our inputs with additional state information (e.g. speed, acceleration, heading angle, vehicle's type, etc) and introduce SSAE to automatically extract features from diverse, inconsistent and ambiguous inputs to form an effective and deep feature representation -- \textit{vehicle descriptor}, which reflects the interaction between various states. 
		
		\item[2.] In order to balance the accuracy of the local interaction position and the perception range of the global traffic scene, we propose a new social pooling method, which introduces a new dilated convolutional network structure to encode the \textit{social vector} in the road occupancy map to extract the spatial interaction features.
	\end{itemize}

	We use the NGSIM US-101 and I-80 datasets to validate our model. The results show that our method is better than the existing benchmark model, which confirms the effectiveness of our method.
	
	\section{Problem Formulation}
	In this paper, we describe the trajectory prediction problem as predicting the probability distribution of the trajectory position of the host vehicle in the future time range based on the historical states information of the predicted vehicle and surrounding vehicles.
	
\begin{figure*}[htbp]
		\centerline{\includegraphics[width=16cm]{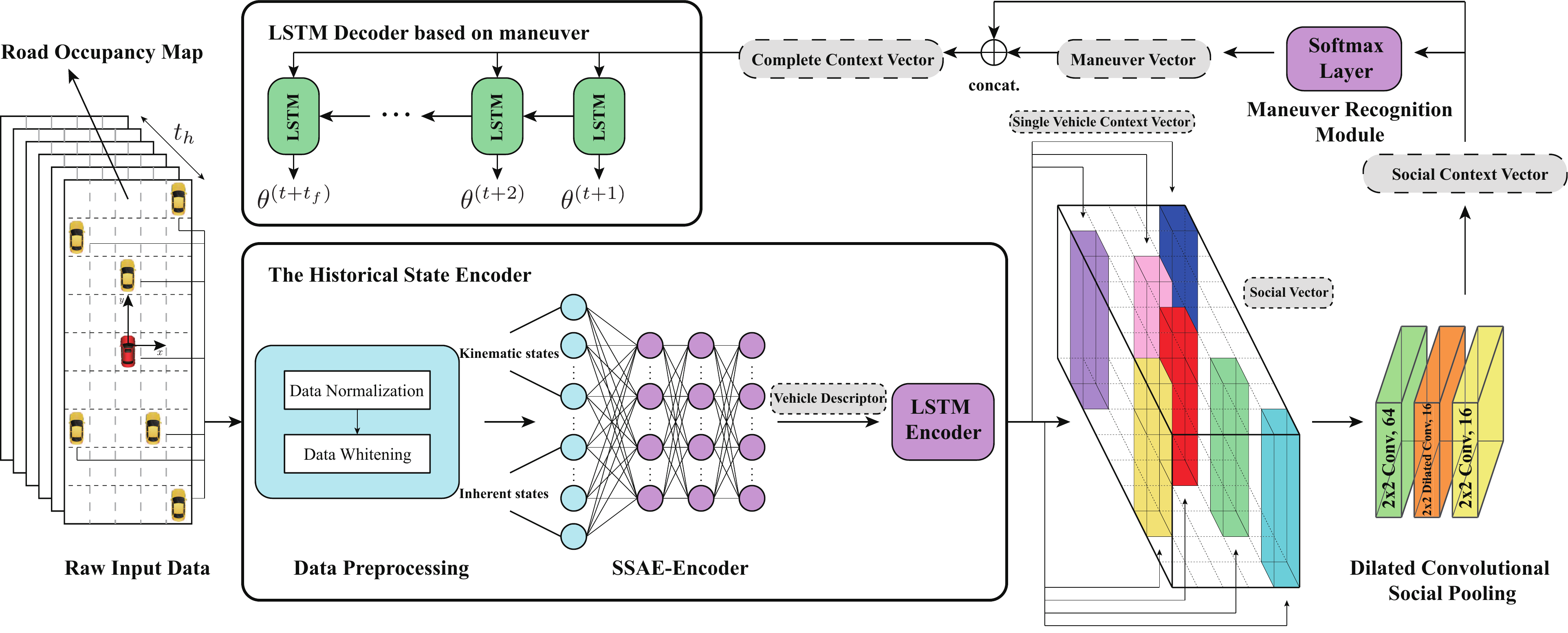}}
		\caption{\textbf{Our Proposed Model}: The \textit{historical state encoder} encodes the historical multi-modal state information and dilated social pooling layers model spatial interactions of all vehicles in the road occupancy map to form a \textit{social context vector}. After the maneuver recognition module, the maneuver vector is concatenated with the \textit{social context vector} to become a \textit{complete context vector} passed into the LSTM decoder. Finally, the LSTM decoder outputs the probability distribution of the trajectory position at each instant $t$ in the future.}
		\label{overall model}
	\end{figure*}

	In order to conveniently represent the vehicle's motion parameters, we use a fixed frame of reference, with the origin fixed at the predicted vehicle at instant $t$, as shown in the Fig.\ref{overall model}. The y-axis points to the direction of motion of the highway and the x-axis is perpendicular to the y-axis.
	For each trajectory, we divide it into 8 parts. The first 3s ($t_{h} = 3s$) represents the historical trajectory and the last 5s ($t_{f} = 5s$) represents the predicted trajectory.
	Finally, the input of the model is expressed as:
	$$
	X = [\mathrm{x}^{(t-t_h)}, ..., \mathrm{x}^{(t-1)}, \mathrm{x}^{(t)}]
	$$
	where,
	$$\mathrm{x}^{(t)}= [s_0^{(t)}, s_1^{(t)},  ..., s_n^{(t)}]$$ 
	
	$ \mathrm{x}^{(t)}$ is the set of states of the predicted vehicle and its surrounding vehicles in the occupancy map at $t$. $s_n ^{(t)}$ represents the multi-modal states of the $ n_{th}$ vehicle at $ t $. $ n $ represents the number of vehicles in the predicted vehicle interactive neighborhood.
	
	The completeness of the model's input greatly affects the accuracy of trajectory prediction. In the past, the vehicle's position coordinates were often used as input, which could not accurately describe the transient state of the vehicle.
	Therefore, we add input information in this article. Corresponding to each time instant $t$ in the historical time-domain $t_{h}$, we input the vehicle's position coordinates, instantaneous speed, acceleration, heading angle, current lane ID and other kinematic states and the inherent static states such as the type and size of the vehicle to describe the transient state $s$ of the vehicle as detailed as possible.
	\begin{equation}\label{s}
	s= [x, y,\Delta x,v,a,\psi,W, L, C,ID]
	\end{equation}
	
\begin{figure}[htbp]
		\centerline{\includegraphics[height=3cm]{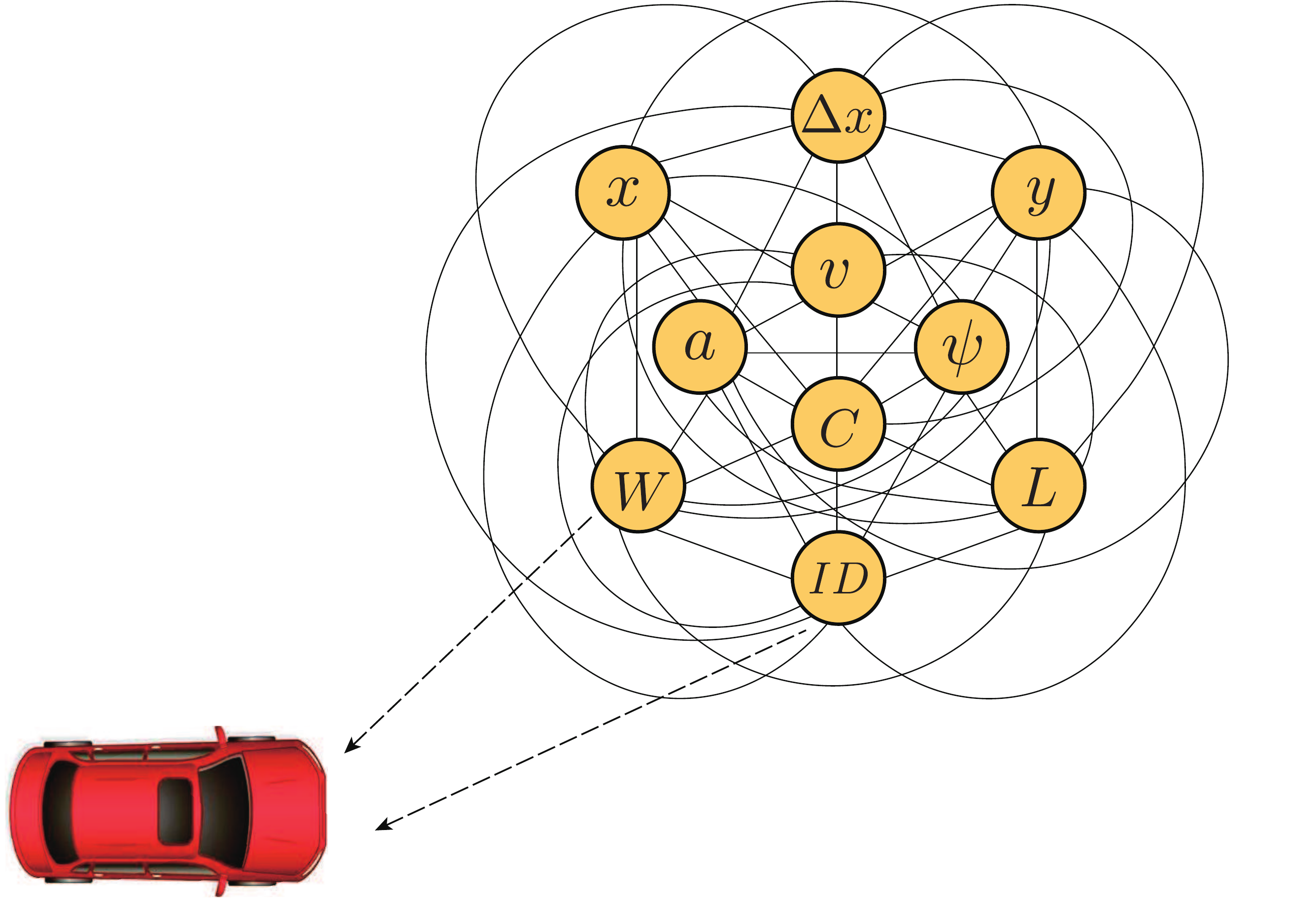}}
		\caption{Multi-modal state input information of the vehicle}
		\label{multimodal_input}
	\end{figure}

	With the increase of vehicle states of the input, we need to consider the interaction of various states in the implicit space. Fig. \ref{multimodal_input} vividly shows the vehicle's multi-modal state information and their interactive relationships contained in the input of this article.
	
	For every input frame of each vehicle, a total of 10 status features are passed into the model and Table \uppercase\expandafter{\romannumeral1} shows the physical meaning of the features of each dimension of the transient state $s$.
	
	\begin{table} 
		\centering
		\caption{Multi-Modal Features For One Input Frame} 
		\begin{threeparttable}
			\begin{tabular}{cc}
				\toprule[1pt]
				Feature& Description\\
				\midrule[1pt]
				$ x $ & Local\; lateral\;coordinate\\
				$ y $ & Local\;longitudinal\;coordinate\\
				$\Delta x$ & Lateral\;distance\;to\;the\;current\;lane\;center\\
				$ v $ & Instantaneous\;velocity\;of\; the vehicle\\
				$ a $ & Instantaneous\; acceleration \;of \; the vehicle\\
				$ \psi $ & Instantaneous\;heading\;angle\;of\;the vehicle\\
				$ W $ & Width\;of\; the vehicle\\
				$ L $ & Length \;of\; the vehicle\\
				$ C $ & Class\;of\; the vehicle\\
				$ ID $ & Current\;lane\;ID\;of\; the vehicle\\
				\bottomrule[1pt]
			\end{tabular}
		\end{threeparttable}
	\end{table}
	Among them, the heading angle of the vehicle can be calculated from the position coordinates (x, y) of the vehicle:
	$$
	\psi =arctan \left( \frac{x^{(t)}-x^{(t-3)}}{y^{(t)}-y^{(t-3)}}\right)
	$$
	
	By entering the transient multi-modal states $s$ for each instant, we provide the model with as comprehensive information as possible for each vehicle.
	
	The model outputs a probability distribution of locations at each frame over the predicted time-domain $t_f$:
	$$
	Y = [\mathrm{y} ^ {(t+1)}, ...,\mathrm{y} ^ {(t + t_f)}]
	$$
	where,
	$$
	\mathrm{y} ^ {(t + 1)} = [x^{(t + 1)}, y^{(t + 1)}]
	$$
	is the future position coordinates of the predicted vehicle at the instant $ t + 1 $.

	\section{Model}
	 Our proposed model consists of a multi-modal state encoding module, an LSTM encoder, dilated convolutional social pooling layers and a maneuver-based LSTM decoder. It is shown in Fig. \ref{overall model}.
	
	\subsection{Multi-modal state encoding module}
	
	\subsubsection{\textbf{Data preprocess}}
	
	For each vehicle, at the instant $t$, the input multi-modal state information is $s$ as given in Equ. \ref{s}.
	It can be seen that the raw input data $s$ has the characteristics of diversity, inconsistency and redundancy, so we need to preprocess it.
	
	In order to make the weight of the impact of the input of each dimension on the objective function consistent, we use the Min-Max normalization \cite{han2011data} to map the data to the range of [0,1] uniformly to get the normalized data $ s^{'} $. 
	Then, in order to remove redundant information in the data and reduce the correlation between each dimension of the input data, we use the ZCA (Zero-phase Component Analysis) whitening transformation on $ s^ {'} $ to get $ s^ {''} $.

	\subsubsection{\textbf{Stacked Sparse Auto-Encoders}}

\begin{figure}[htbp]
		\centerline{\includegraphics[height=6cm]{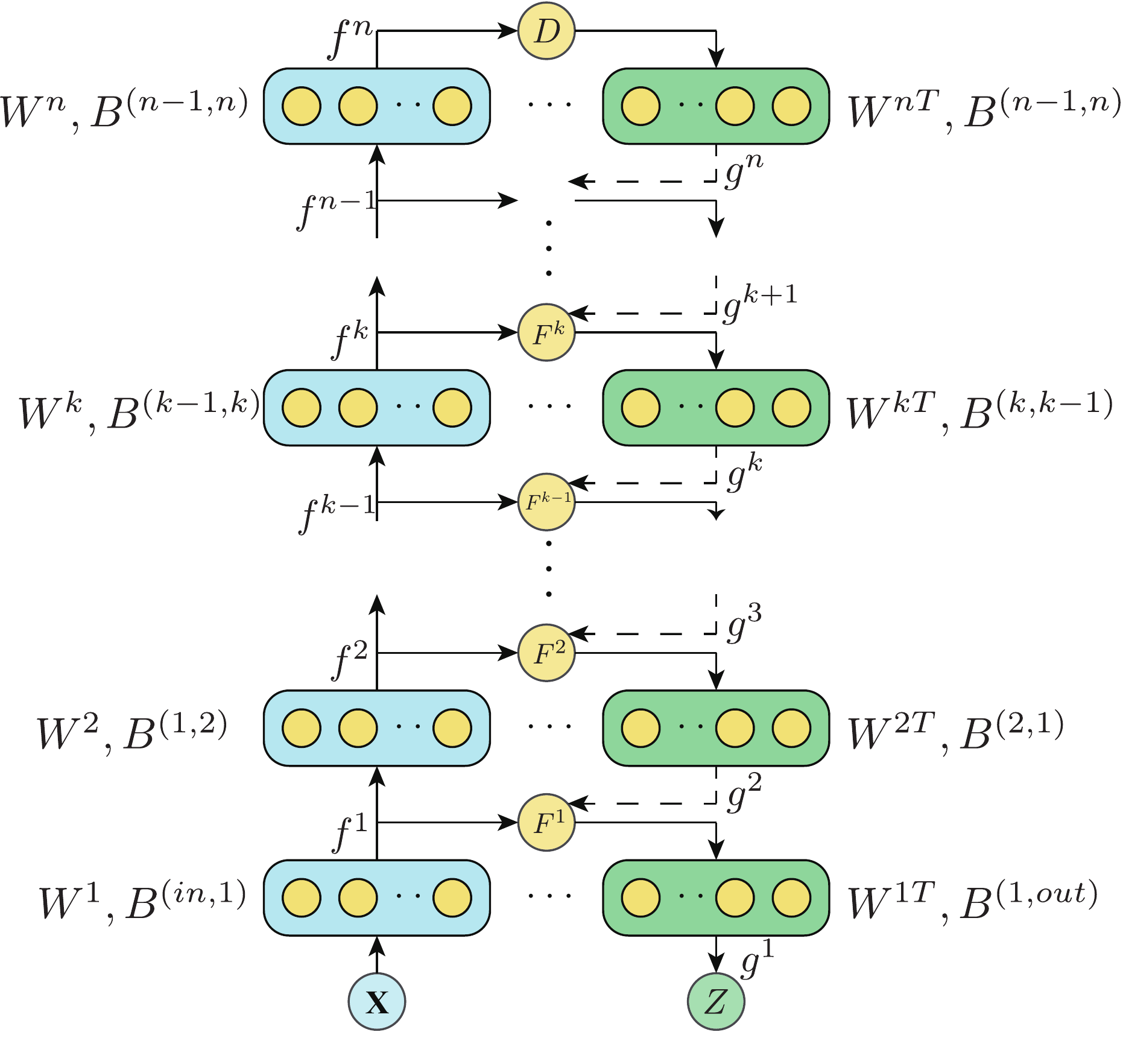}}
		\caption{Structure of the SSAE Model}
		\label{SSAE}
	\end{figure}

	The features of the input data significantly affect the performance of trajectory prediction. The past methods usually only take the vehicle's position coordinates as the input of the model and rely on artificial design to obtain features. However, in this article, we input complex multi-modal high-dimensional data and cannot obtain deep features through artificial means. Therefore, we introduce a deep learning based model—a stacked sparse auto-encoder (SSAE) to perform unsupervised feature extraction on the whitened data $ s^ {''} $. SSAE is a deep neural network composed of several auto-encoder structural units with sparse representation constraints, as shown in Fig.\ref{SSAE}.
	
	The auto-encoder includes two processes of encoding and decoding. Firstly, the input data obtains a new feature representation through the encoder and then the new feature representation is processed by the decoder to obtain the output. After each complete processing, the error must be back-propagated and continuously optimized, so that the output and input are as similar as possible. The increase in the number of layers of the auto-encoder can make the feature representation learned from the original data deeper. Adding the sparse representation constraints can enable the auto-encoder to get more compact data expression and improve the generalization ability of the model.
	
	During training, given a data set of $ N $ samples $ S ^ {''} = [s_ {1} ^ {''}, s_ {2} ^ {''}, ..., s_ {i} ^ {''}, ..., s_ {N} ^ {''}] $, for each sample $ s_ {i} ^ {''} $ in $S ^ {''}$, we can obtain a corresponding deep feature representation — \textit{vehicle descriptor} $ {d_i} $ trough the encoder of the SSAE model, therefore a new feature encoding set $ D = [d_1, d_2, ..., d_i, ..., d_N] $ is formed.
	We take the first layer of the SSAE model as an example to illustrate the encoding and decoding process.
	
	\begin{equation}\label{SSAE_encoder}
	F^1 = f^1(W^1 S^{''} + B^{(in,1)}) 
	\end{equation}
	
	\begin{equation}\label{SSAE_decoder}
	Z = g^1(W^{1T}F^1 + B^{(1,out)})\approx S^{''}
	\end{equation}
	
	where, $S^{''}\in R^{n}$ (n is the number of neurons in the input layer) represents a set of input samples of the encoder, $ F^1 \notin R ^ {n} $ represents a set of hidden feature expressions for the decoder's reconstructed samples and $Z$ is the set of reconstructed samples; $ f ^ 1 $ and $ g ^ 1 $ respectively represent the activation function of the input layer and the hidden layer, which are relu and linear functions respectively; $W^{1}$ and $B^{(in,1)} $ are the weight matrix and bias vector from input layer to hidden layer respectively; $ W^{1T}$ and $B^{(1,out)} $ are the corresponding coefficient from hidden layer to output layer.
	Similarly, for the $k_{th}$ layer, there are:
	
	\begin{equation}
	F^k = f^k(W^k F^{k-1} + B^{(k-1,k)}) 
	\end{equation}
	
	\begin{equation}
	Z^{k-1} = g^k(W^{kT}F^k + B^{(k,k-1)})\approx F^{k-1}
	\end{equation}
	
	By adjusting the parameter set $\Phi$ of the SSAE model, we want to make $S^{''}$ and $Z$ as approximate as possible. After adding the sparse representation constraints, the mathematical expression of our loss function is as follows:
	\begin{equation}\label{SSAE_loss}
	J_{sparse}(\Phi) = L(S^{''}, Z) +\mu\rho(\Phi)+\lambda||\Phi||^2 
	\end{equation}
	
	Among them, $ L (S ^ {''},R) $ is the reconstruction error expressed in MSE form and $ || \Phi || ^ 2 $ is regularized term to avoid overfitting. $\mu $ and $ \lambda $ are the weight coefficients for the sparse penalty term and the regular term. $ \rho (\Phi) $ is the sparse penalty term, given as:
	 
	\begin{equation}
	\rho_j = \frac{1}{N}\sum_{i=1}^{N}d_{i,j}
	\end{equation}
	
	\begin{equation}
	KL(\rho || \rho_{j})=\rho log\frac{\rho}{\rho_j}+(1-\rho)log\frac{1-\rho}{1-\rho_j}
	\end{equation}
	
	\begin{equation}
	\rho(\Phi)=\sum_{j=1}^{n}KL(\rho || \rho_{j})\\\\
	\end{equation}
	
	where, $ N $ is the number of training samples, $i$ represents the $i_{th}$ sample in $N$ samples, $\rho_j$ represents the average activation probability of the $ j_{th} $ neuron in the hidden layer, $ \rho $ is a hyper parameter with a particularly small preset value, $n$ is the number of hidden neurons,

	By minimizing the loss function as given in Equ. \ref{SSAE_loss},
	we can reduce the reconstruction error and obtain the sparse feature representation — the \textit{vehicle descriptor} to represent the deep and high-level features of the original multi-modal input.
	
	In this paper, we use the encoder part of the SSAEs model on the multi-modal state input of each vehicle at any time instant for feature selection and extraction to generate a vehicle descriptor $ d $ , which will be passed to the next steps.
	
	\subsection{LSTM encoder}
	Now, we have obtained the vehicle descriptor $d$ of each vehicle at any time instant $t$. In this section, we will arrange the \textit{vehicle descriptor} $d$ corresponding to each time instant in the historical time-domain in chronological order to form time sequence information $H$ for each vehicle.
	
	\begin{equation}
	H = [{d}^{(t-t_h)}, ..., {d}^{(t-1)}, {d}^{(t)}]
	\end{equation}
	
	LSTM networks are good at dealing with time sequences, so we use LSTM networks to encode the input historical information and mine the memory characteristics of trajectory sequences $H$.
	At each time instant, the LSTM unit reads the vehicle descriptor $ d ^ {t} $ input at the current time instant $t$ and the hidden state $ h ^ {(t-1)} $ passed at the previous time to update the current hidden state $ h ^ {(t)} $, that means, $ h ^ {(t)} = f (h ^ {(t-1)}, d ^ {t}) $. In this way, the LSTM network learns the rules from the historical trajectory sequence and encodes the historical information of each vehicle into a fixed-length context vector, which reflects the understanding and memory of the historical trajectory of the vehicle by the LSTM encoder. We record this vector describing the historical multi-modal states of each vehicle as \textit{single vehicle context vector}, as shown in Fig. \ref{overall model}.
	
	\subsection{Dilated convolutional social pooling}
	In the common traffic scene, the motion of each vehicle will affect the decisions of surrounding vehicles and all the vehicles in the scene together form an interdependent, dynamically balanced whole. Therefore, in order to obtain better prediction accuracy, our model needs to capture the interactive relationships between vehicles and dynamically adjust the trajectory of the predicted vehicle according to the movement of surrounding vehicles.
	
	In order to model the interaction between vehicles, we build a road occupancy map based on the vehicle occupancy of the road. Due to the specification of the highway structure, the road occupancy map is composed of block units of the same size. The length of the block area is the average length (15 feet) of an ordinary car and the width is set to the width of the lane. In the horizontal aspect, we consider the predicted vehicle's lane and the four nearest left and right lanes; in the longitudinal aspect, we use the predicted vehicle as to the center and select 4 road occupancy units (60 feet) forward and backward. Consequently, a road occupancy map consisting of 9$\times$5 rectangular block units is formed as shown in Fig.\ref{overall model}.
	Then, the \textit{single vehicle context vector} of each vehicle in the neighborhood will fill this map according to the spatial position in the traffic scene to form a \textit{social vector}. Fig.\ref{overall model} gives a \textit{social vector} visually. In this way, the historical state information and spatial position of surrounding vehicles are introduced into our model.
	
	To a certain extent, the map is similar to an image, each road block unit is equivalent to a pixel and the \textit{single vehicle context vector} at the corresponding position is equivalent to the pixel value, so we can introduce common processing mechanism for image information.
	To learn the spatial interaction between vehicles in the scene, we use convolutional layers to process the \textit{social vector}. Convolutional Social Pooling \cite {cs-lstm} first introduced convolutional layers and max-pooling layers on the \textit{social vector} containing the historical information of the neighborhood vehicle. However, the original pooling layer will cause the loss of local position information and reduce the resolution. 
	
	For the trajectory prediction task, we need to capture not only global information reflecting the driving scene but also accurate local interaction position information. We want networks to have a larger field of view so that they can \textit{see} larger areas to make better decisions. At the same time, we hope that the network can focus on the local interactive location and reduce the loss of local accurate location features. 

\begin{figure}[htbp]
		\centerline{\includegraphics[height=3cm]{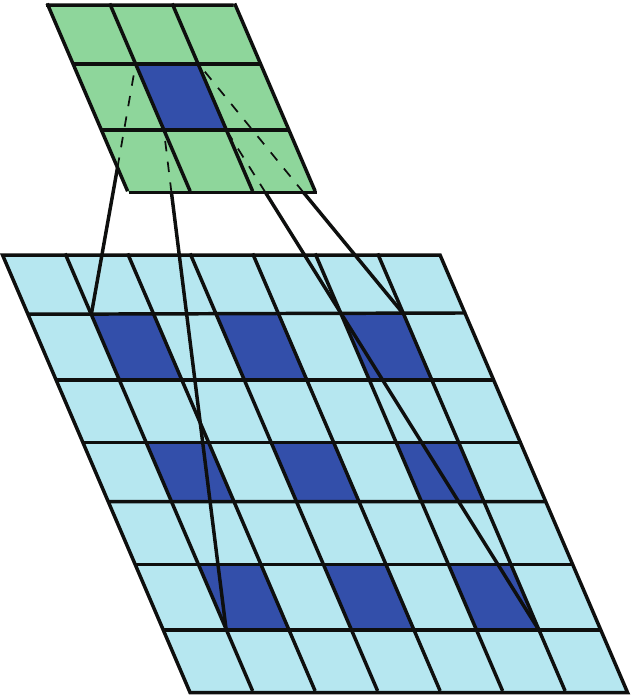}}
		\caption{Dilated-Convolution with a 3$\times$3 kernel and dilation rate 2}
		\label{dilated_convolution}
	\end{figure}

	Therefore in this paper, we abandoned the pooling layer and designed a novel dilated convolutional social pooling architecture that extracts interactive features from the \textit{social vector} to form the \textit{social context vector}. A typical working process with dilated convolution is shown in Fig. \ref{dilated_convolution}. Our social pooling layer is composed of three convolutional layers as shown in Fig.\ref{overall model}. By passing the \textit{social context vector} to the next component, the LSTM decoder can utilize useful interactions from it and predict the future position of the vehicle.
	
	\subsection{Maneuver-based LSTM decoder}
	Due to the complexity of traffic scenes and the diversity of driving styles, the driving trajectories of traffic participants in actual scenes are often uncertain. 
	To represent the inherent uncertainty of the future trajectory, we use an LSTM decoder to generate a probability distribution of vehicle positions in the next $ t_f $ time and five parameters of the binary Gaussian distribution are used as the output of the decoder. 
	
	In addition, the generation of the actual trajectory is closely related to the driving maneuver. If the driver's maneuver is not judged when making trajectory prediction, the output result tends to the average of different maneuver trajectories, which obviously does not conform to the actual scenario. 
	Therefore, we introduce a maneuver recognition module to obtain the driver's maneuvers to assist in trajectory prediction.
	
	In the highway driving scenario, there are three common driving maneuvers: lanes keeping, changing lanes to the left and right. 
	In this article, we take the \textit{social context vector} as input $I$ and use the $ \mathrm {Softmax} $ function to calculate the probabilities of three driving maneuvers $P(m_i|I)$. Among them, $ m_i (i = 1,2,3) $ respectively represents three types of maneuvers. We specify the category with the highest probability as the correct maneuver and predict the probability distribution of the future trajectory based on this maneuver. The \textit{maneuver vector} $ M $ is a one-hot vector output from the maneuver recognition module.
	In order to predict the future trajectory based on the maneuver, $ M $ is combined with the \textit{social context vector} to form a \textit{complete context vector} as the input of the LSTM decoder.
	
	Consequently, our model estimates the probability distribution $ P (O|I) $ of future trajectories based on maneuver.
	\begin{equation}\label{distribution}
	P(O|I) = \sum P_ {\theta} (O | \, m_i, I) P (m_i | I)
	\end{equation}
	where,
	\begin{equation}
	\theta = [\theta^{(t + 1)}, ..., \theta ^{(t + t_f)}]
	\end{equation}
	corresponds to the parameters of the binary Gaussian distribution of the predicted trajectory positions at each instant in the future. It includes the mean and variance of the position coordinates, reflecting the uncertainty of the predicted trajectory.
	Finally, through training the model, we can determine the parameters of the binary Gaussian distribution.
	
	\subsection{Implementation details}
	
	In order to reduce the computational cost, we set the downsampling rate to 2 over the NGSIM Dataset. We use 80\% of the entire dataset as the training set and 20\% as the test set. 
	Through end-to-end training, we want to minimize the negative log likelihood loss as given in Equ. \ref{NLL} across all training data points.
	\begin{equation}\label{NLL}
	Loss =-log\left(\sum_{i}P_{\theta}(O|\,m_i,I)P(m_i| I)\right)
	\end{equation}
	
	We use Adam \cite{kingma2014adam} to train the model and set the learning rate as 0.001. The LSTM encoder has a 64-dimensional state and the decoder has a 128-dimensional state. The sizes of the dilated convolutional social pooling layer are shown in Fig.\ref{overall model}. We use $ \alpha = 0.1 $ for all layers activated by leaky-ReLU. The model is implemented using $ PyTorch $ \cite{pytorch}.
	We use the RMSE of the predicted trajectory and the true future trajectory to evaluate our complete model, as done in \cite{Gail-gru}.
	
	\section{Experiment and Results}
	
	\subsection{Compared models}
	We report the RMSE of the compared models listed in the Table \uppercase\expandafter{\romannumeral2} over the prediction time range $ t_f $ as done in \cite{Gail-gru}.
	In this paper, we establish a set of comparative models base on our model to verify the important role of each component of our model as follows.
	\begin{itemize}
		\item DCS-LSTM: LSTM with dilated convolutional social pooling.
		\item K-Model: DCS-LSTM with kinematic inputs s = [x, y, $\Delta x $, v, a, $\psi$, ID].
		\item MM--Model: DCS-LSTM with multi-modal state inputs s = [x, y, $\Delta x$, v, a, $\psi$, ID, W, L, C].
		\item VD+DCS-LSTM (our proposed model): LSTM with the \textit{vehicle descriptor} and dilated convolutional social pooling.
	\end{itemize}

	\begin{table} 
		\centering
		\caption{Performance Comparison with baseline models} 
		\begin{threeparttable}
			\begin{tabular}{cccccc}
				\toprule[1pt]
				Model& 1s& 2s& 3s& 4s& 5s\\
				\midrule[1pt]
				CV \cite{cs-lstm} & 0.73& 1.78& 3.13&  4.78&  6.68\\
				C-VGMM + VIM \cite{deo2018would} & 0.66& 1.56& 2.75&  4.24&  5.99\\
				GAIL-GRU \cite{Gail-gru}& 0.69& 1.51& 2.55&  3.65&  4.71\\
				S-LSTM \cite{social-lstm}&  0.65& 1.31& 2.16&  3.25&  4.55\\
				M-LSTM \cite{multi-modal}& 0.58& 1.26& 2.12&  3.24&  4.66\\
				\textbf{VD+DCS-LSTM} & \textbf{0.57}& \textbf{1.22}& \textbf{2.01}& \textbf{2.99}& \textbf{4.23}\\
				\bottomrule[1pt]
			\end{tabular}
		\end{threeparttable}
	\end{table}
	
	\subsection{Results}
	
\begin{figure}[htbp]
		\centerline{\includegraphics[width=9cm]{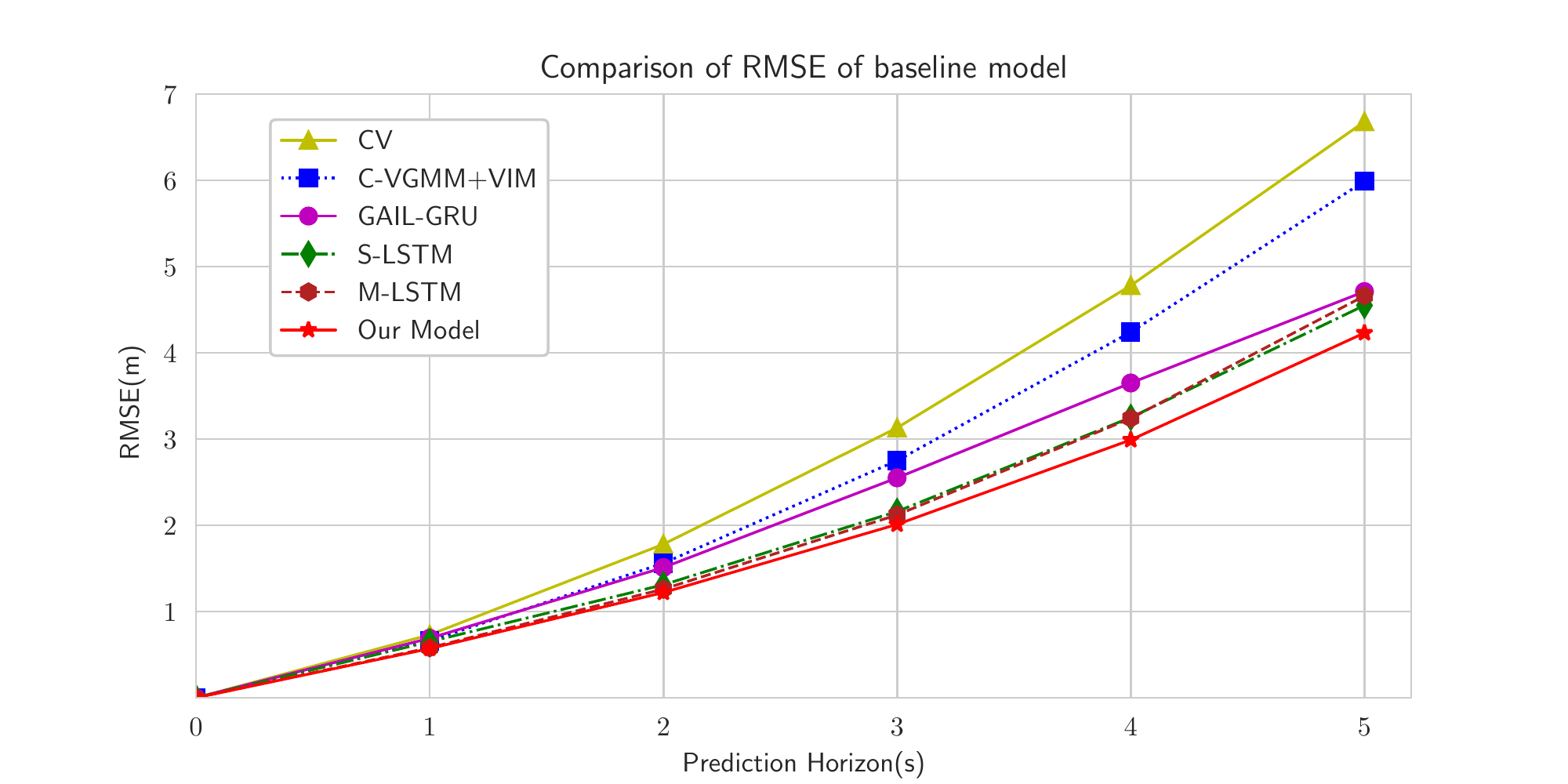}}
		\caption{Comparison of Quantitative Results of Different Models}
		\label{baseline}
	\end{figure}

	Table \uppercase\expandafter{\romannumeral2} and Fig.\ref{baseline} shows the values of RMSE between the predicted and actual trajectories of all baselines.
	From the quantitative results in the Table \uppercase\expandafter{\romannumeral2}, we can see that our model performs better than all other baseline models, proving our model's effectiveness.
	
	We can see that the prediction errors of the CV baseline and the C-VGMM + VIM model are obviously higher than other models, which shows that the recurrent neural network has a strong advantage in dealing with highly nonlinear vehicle trajectory sequence problems and is good at excavating time sequences data deeply, especially shows excellent advantages in long-term prediction.
	
	In addition, we can note that the performance of S-LSTM, M-LSTM and our model is better than V-LSTM. This suggests that considering the motion of surrounding vehicles of the predicted vehicle in the model is very critical to improve the prediction accuracy.
	
	Finally, we notice that the error values of M-LSTM and S-LSTM are close. This is because the M-LSTM uses the predicted trajectory based on the maneuver of the highest probability to calculate the value of the RMSE, but the maneuver recognition module may misclassify maneuvers, so the error value of M-LSTM is slightly larger at some predicted instants. But the trajectory predicted based on maneuver fits better with the real trajectory. Moreover, the correct driver's maneuver can assist the supervision of the trajectory prediction task and improve the prediction accuracy.
	
	\subsection{Ablative analysis}
	In order to study the relative significance of each component in the model for trajectory prediction, we performed an ablative analysis on each component of the model. In particular, we focus on the importance of the two key components of the vehicle descriptor (VD) and dilated convolutional social pooling (DCS). Table \uppercase\expandafter{\romannumeral3} shows the root mean square value of the prediction error of each ablative model.

	\textbf{1. Dilated Convolutional Social Pooling}
	
	By comparing the predictive errors of the S-LSTM and CS-LSTM models, we observe that the CS-LSTM model performs better than the S-LSTM model. These two models represent two different ways of modeling vehicles' interaction. Through experimental results, we can know that compared to the fully connected social pooling, the convolutional social pooling can better extract the interactive features, which is because the latter method can preserve the spatial relationship between vehicles.
	
	By comparing CS-LSTM and DCS-LSTM, we find that DCS-LSTM gets better predictive performance. This also shows that the dilated convolutional network can extract the interactive features of vehicles more deeply. The max-pooling layer proposed in \cite{cs-lstm} loses precise local position information during dimension reduction. However, the dilated convolutional network can ensure that the same dimension reduction effect is achieved while retaining more accurate local position information. Therefore, the dilated network we use in this article plays a positive role in trajectory prediction.
	In this paper, we use dilated convolutional social pooling to increase the receptive field of the network without sacrificing \textit{resolution}. In this way, we can better solve the mutual exclusion problem of global scene receptive field and local position's resolution, so as to better capture the interactive features between vehicles and improve the accuracy of trajectory prediction.
	
	\textbf{2. Vehicle Descriptor}
	
	\begin{table} 
		\centering
		\caption{Performance Comparison with Ablative models} 
		\begin{threeparttable}
			\begin{tabular}{cccccc}
				\toprule[1pt]
				Model& 1s& 2s& 3s& 4s& 5s\\
				\midrule[1pt]
				V-LSTM \cite{cs-lstm}& 0.68& 1.65& 2.91&  4.46&  6.27\\
				S-LSTM  \cite{social-lstm}& 0.65& 1.31& 2.16&  3.25&  4.55\\
				CS-LSTM  \cite{cs-lstm}& 0.61& 1.27& 2.09& 3.10&4.37\\
				DCS-LSTM & 0.60& 1.26& 2.06& 3.07&  4.33\\
				MM-Model  & 0.67& 1.48& 2.46& 3.51& 4.73 \\
				K-model  & 0.58& 1.25& 2.03& 3.04&  4.29\\
				\textbf{VD+DCS-LSTM} & \textbf{0.57}& \textbf{1.22}& \textbf{2.01}& \textbf{2.99}& \textbf{4.23}\\
				\bottomrule[1pt]
			\end{tabular}
		\end{threeparttable}
	\end{table}

	By comparing the K-Model and DCS-LSTM model, it can be seen that K-Model can obtain smaller prediction errors. This also accords with our intuitive feelings. When our input is more detailed, we can better describe the state of the vehicle at each moment. This will help improve our prediction accuracy.
	
	However, by comparing the K-Model and MM-Model models, we can observe that although the modalities of the input information have increased and the input has become more detailed, the model's predictive performance has not become better and even worse than the model that only simply inputs the historical trajectory coordinates of the vehicle.
	This result indicates that the deep features of the transient state of the vehicle cannot be extracted from the multi-modal input composed of different types of state information in a simple way.
	
	Through the previous comparison, we notice that adding kinematic information such as vehicle's speed, acceleration, heading angle and distance from the centerline based on the input trajectory position coordinates information can improve the model's predictive accuracy.
	This is because the current multi-modal input belongs to the kinematic information of the vehicle and has a similar type to a certain extent. Therefore, when the kinematics information of the vehicle is used as a whole input, the model can learn how to better describe the kinematics state of the vehicle at each instant. The input information of the MM-Model no longer only includes the kinematic information of the vehicle, but also adds the inherent properties of the vehicle such as the vehicle's type and size. In the case of this type of input, we need to consider the problem of how to interact deeply between the states of each dimension from the multi-modal input with inconsistent types and different semantics states and affect the trajectory prediction together. Therefore, the MM-Model adds input information but does not obtain a good prediction result that can be explained.
	
	In the end, by comparing all models, we noticed that the VD+DCS-LSTM model had the smallest error over the entire predictive range and reached the most advanced level.
	
	Compared with the MM-Model, which directly inputs multi-modal information without any processing, we introduce the SSAE in the VD+DCS-LSTM model to encode the multi-modal input information of the vehicle, effectively extracting the deep interactive features between vehicle's different states. By using the \textit{vehicle descriptor} encoded by SSAE to characterize the state of the vehicle at each moment.
	
	SSAE automatically learns and sorts out the interaction relationship between different dimensional states of the vehicle's multi-modal input information. The multi-modal input information is used to the greatest extent, so the predictive effect is improved.
	Especially with the increase of the predictive time range, our model demonstrates its ability to control errors and achieves reliable prediction of future positions over a longer time range. The state of the vehicle at each moment is more fully described by the \textit{vehicle descriptor}, which helps us to understand the vehicle's past behavior more deeply and obtain more accurate maneuver classification results, consequently improving the prediction accuracy.
	It can be seen from the comparison results that our vehicle descriptors have played a significant role in the process of error control, because the vehicle descriptors can modify the results of the maneuver recognition module and further assist in monitoring the trajectory predictive results.
	
	\section{Conclusions and Future work}
	This paper presents a new trajectory prediction model with the \textit{vehicle descriptor} and dilated convolutional social pooling. The results show that we outperform state-of-the-art methods on the two public vehicle trajectory datasets, proving the validity of our overall model. The ablative analysis illustrates that our model makes full use of the vehicle's multi-modal state information by using the \textit{vehicle descriptor} to capture deep features and extracts interactive features with dilated convolutional social pooling.
	
	A limitation of our method is that it corresponds to the highway scene and the input of multi-modal information only consists of vehicle states. Future work can be focused on extending to other scenarios and adding the driver's own factors to the multi-modal information.

	\bibliographystyle{ieeetr}
	\bibliography{ref}
	
\end{document}